# Wakefield issue and its impact on X-ray photon pulse in the SXFEL test facility

Minghao Song[1,2], Kai Li[1,2], Chao Feng[1], Haixiao Deng[1,*], Bo Liu[1], and Dong Wang[1]

*1 Shanghai Institute of Applied Physics, Chinese Academy of Sciences, Shanghai, 201800, P. R. China*

*2 University of Chinese Academy of Sciences, Beijing 100049, P. R. China*

**Abstract**：

Besides the designed beam acceleration, the energy of electrons changed by the longitudinal wakefields in a real free-electron laser (FEL) facility, which may degrade FEL performances from the theoretical expectation. In this paper, with the help of simulation codes, the wakefields induced beam energy loss in the sophisticated undulator section is calculated for Shanghai soft X-ray FEL, which is a two-stage seeded FEL test facility. While the 1$^{st}$ stage 44 nm FEL output is almost not affected by the wakefields, it is found that a beam energy loss about 0.8 MeV degrades the peak brightness of the 2$^{nd}$ stage 8.8 nm FEL by a factor of 1.6, which however can be compensated by a magnetic field fine tuning of each undulator segment.

*Corresponding author: denghaixiao@sinap.ac.cn

1. Introduction

Free-electron laser (FEL) at X-ray wavelengths delivers intense pulses on ultra-short time scales, which is used to detect the dynamic processes, such as chemical-bond formation, charge transfer and light-induced superconductivity, or to characterize the macromolecular structure without damage [1, 2]. Currently, several X-ray FEL facilities are under operation and/or construction around the world [3-7]. In pursuit of fully coherent FEL pulses at 8.8 nm wavelength, the baseline design of Shanghai soft X-ray FEL (SXFEL) test facility is two-stage seeded FEL scheme [8].

As well known, stringent beam energy control and fine undulator magnetic field set along the whole undulator system is required to maintain the FEL lasing and the FEL bandwidth. In a real FEL machine, the electron beam is not always on the targeted energy because of the drift and instability of the accelerating field and the beam energy loss due to the longitudinal wakefields [9]. In general, the beam energy deviations in the LINAC can be monitored and compensated by the beam energy feedback system before entering the undulator sections. While it is the passive wakefields within the undulator section that degrade the FEL performance by pushing electrons off their resonance energies, especially within those sophisticated vacuum chambers and pipes for a multi-stage seeded FELs. Therefore in this paper, on the basis of technical design aspects of SXFEL, the wakefields caused beam energy loss, its impact on X-ray FEL pulse generation, and the coping strategy are presented. A distributed wakefields arisen from resistive wall [10-12], surface roughness [13-19] and discontinuities of beam pipes [20-23] are considered. It shows that, the peak brightness of the final 8.8 nm FEL will degrade by a factor of 1.6 with a gradual beam energy drop off about 0.8 MeV in the whole undulator. It can be compensated by a magnetic field fine tuning of the radiator undulator.

This paper is organized as follows. Firstly, an introduction of Shanghai soft X-ray FEL test facility is briefly described. In the following section, numerical calculations on the resistive wall, surface roughness and geometrics wakefields are given. Furthermore, the beam energy loss impact on FEL pulses and the coping strategy are illustrated. Finally, we summarize the results of the paper.

## 2. Introduction of SXFEL

The SXFEL is based on the frequency up-conversion scheme of an initial coherent seed pulse in an FEL amplifier employing multiple undulators, namely high-gain harmonic generation (HGHG) [24-26]. The initial signal is provided by a conventional pulsed laser operating at 264 nm wavelength. The energy of the electron beam is modulated via the resonant interaction with the 264 nm seed laser in the so-called "modulator", then a chromatic dispersive chicane is used to develop density bunching with harmonic components, finally coherent FEL radiation at the 44 nm wavelength, i.e., the 6$^{th}$ harmonic of the seed laser is produced by the micro-bunched beam in the downstream "radiator". In order to achieve the designed wavelength, a second stage HGHG from 44 nm to 8.8 nm is used in SXFEL. As illustrated in Fig. 1, an extra pair of modulator M0 and dispersive chicane DS0 is reserved before the first stage HGHG to enable the echo-enhanced harmonic generation (EEHG) scheme [27, 28], which is expected to overcome the limited frequency up-conversion efficiency of HGHG.

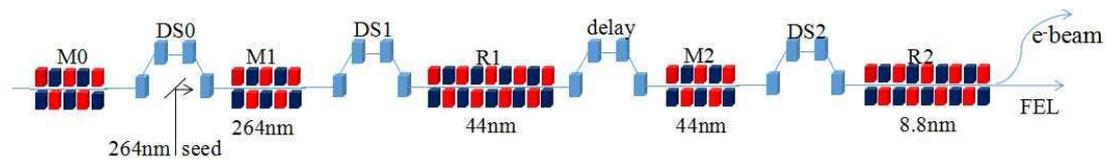

Fig. 1. Schematic layout of SXFEL test facility.

The SXFEL photo-injector is based on the 1.6-cell electron gun developed at BNL/SLAC/UCLA. Following standard layout, the design includes a solenoid for emittance compensation and acceleration to 130 MeV with two S-band sections [29]. The electrons are then sent into the laser heater system where a 795 nm Ti-sapphire laser with the pulse length of 10 ps, to suppress the micro-bunching instability and control the deviation and the distribution shape of sliced beam energy. The main LINAC accelerates the electron beam to 840 MeV and compresses the beam to its final duration and peak current. At the exit of LINAC, depending on the FEL lasing requirements, a bunch length of 1 ps (FWHM) and a peak current of 500 A or higher can be delivered with a 500 pC bunch charge, and the normalized emittance should not exceed 1.0 μm-rad to satisfy the desired photon throughput. These specifications are numerically predicted by the beam dynamic simulations, includes a safety margin against collective instability effects. Meanwhile, the LINAC magnetic focusing system is designed to minimize the emittance dilution due to transverse wakefields, momentum dispersion and coherent synchrotron radiation in bending magnets.

**3. Wake Potential Calculations**

It is worth stressing that the linear accelerator section consists of relative simple and large aperture structures when compared with the sophisticated undulator sections in a seeded FEL facility, and the peak current of the electron beam is low when passing through the impendence items before the final bunch compressor. Therefore, the beam energy loss due to wakefields is a small amount in the linear accelerator, and more importantly it can be compensated by tuning the amplitude and phase of the RF cavities. Therefore in this paper, we concentrate on the wakefields of the undulator sections for SXFEL.

*3.1. The Resistive Wall and Surface Roughness*

The resistive wall and surface roughness are usually considered for the vacuum chambers of the undulator. All the undulators of SXFEL have been chosen to be out vacuum planar undulator, hybrid permanent magnets type. The wavelength can be tuned by changing the undulator gap at constant beam energy. For example, the magnetic length of the individual segment is 3.0 m (containing 75 periods) for the stage-1 radiators and 3.0 m (128 periods) for the stage-2 radiators, respectively. And a 3.34 m long aluminum vacuum chamber with an elliptical cross section of 6×15 mm aperture will be used for all the radiator undualtors, while the stage-1 and stage-2 radiators consist of 3 and 6 undulator segments.

With a round pipe approximation of 3 mm radius, the calculation of the short range wakefields of a resistance chamber can be accomplished by the formulas from ref. [12, 17]. The surface measurements of the vacuum chamber sample by using atomic force microscope [16], similar to that used in SXFEL, agrees well with the small-angle approximation theory [17, 18]. Thus, the roughness model used consists of small amplitude of 100 nm, shallow sinusoidal corrugations.

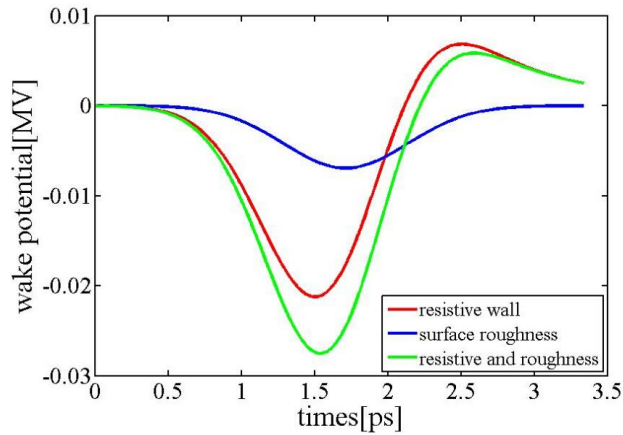

Fig. 2. Wake potential of a 3.34 m long vacuum chamber that has resistive wall (red), roughness (blue), and both resistive wall and roughness (green) .The resistive wall calculation includes AC conductivity

of aluminum.

In Fig. 2, the wake potentials for one segment of the undulator vacuum chamber are presented. It is found that the total wakefields is dominated by the resistive wall effect. Using the results shown in Fig. 2, the beam energy loss within a 3.34 m long undulator chamber can be obtained. For a Gaussian current distribution here, the energy loss due to the resistive wall and roughness are approximately 11.3 keV and 4.9 keV, respectively.

*3.2. Geometrics Wakefield*

In a multi-stage seeded FEL facility, on one hand, the delay chicane and various beam diagnostic devices are placed between different stages. On the other hand, cavity beam position monitors, profiles, quadrupoles, correctors and phase-shifters are installed between the undulator segments to monitor and correct the electron's behaviors. Thus, from the point view of wakefields, within the undulator section, the beam pipes need to be interrupted and connected by flanges, bellows and so on.

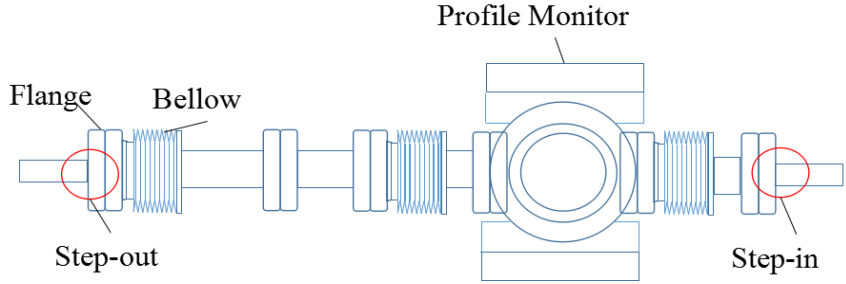

Fig. 3. One module of the insert components between two adjacent undulator segments

Besides the resistance and roughness wakefields mentioned above, here the wakefields generated in these geometrics are also calculated for SXFEL. The Napoly indirect method is employed by the wakefield code ABCI [30] for 2D calculations. For short bunches, ABCI employs a moving mesh that encloses the longitudinal and transverse direction. As a typical example, the insert component module between two adjacent undulator segments (see Fig. 3) is illustrated here, which consists of flanges, bellows, profile monitor, step-out and step-in. From the ABCI simulation, the energy loss of electron beam is 0.4 keV, 5.1 keV, 3.3 keV, 38.5 keV and 0.6 keV caused by the flange, bellows, discontinuity in the profile monitor, and the aperture change of the vacuum chamber(step-out and step-in), respectively.

It is found that the energy loss from flange and step-in is so small that can be neglected. Although the energy loss from bellows and profile monitor is one order of magnitude lager than those from flange or step-in, they are still small when compared to step-out. There is an effort under way to diminish the wake effects of bellows using shielding techniques for SXFEL. Fig. 4 shows the wake potentials of one insert module for comparison study. The beam energy loss is 59 keV and 64 keV for the bellows with and without shielding, respectively. This discrepancy demonstrates that bellows play an appreciable role in SXFEL where the peak current of the electron beam is several hundred Ampere.

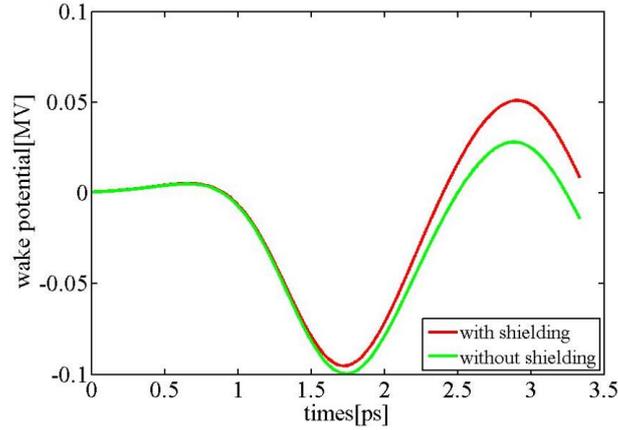

Fig. 4. Wake potential of the inserting module, using bellows with (red) and without (green) shielding.

*3.3. The total energy loss*

In order to study the total energy loss when the electron beam passes through the undulator section, the undulator section is divided into several modules and calculated. In general, the resistive wall and roughness contributions of the undulator chamber are theoretically estimated by wakefield theories and the geometrics between the undulators are numerically modeled by ABCI [30]. Fig. 5 shows the evolution of the wake potential experienced by the electron beam at the different location of the undulator section. Accordingly, the wake effects become more and more serious along the undulator section, and the total energy loss is approximately 0.8 MeV at the end of undulator section.

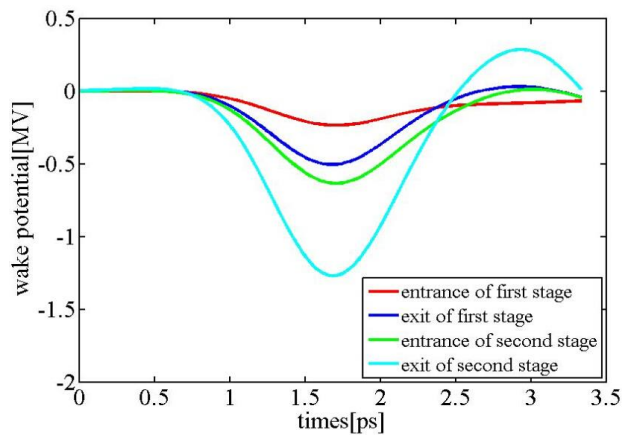

Fig. 5. Wake potentials the electron beam experienced during the undulator section.

**4. Wakefield Impact on FEL Pulse**

In order to assess the wakefields impact on the output photon pulses, the FEL performance has been simulated by the well-benchmarked FEL code GENESIS [31]. A 264 nm laser with 200 MW peak power and 100 fs FWHM duration is used as the initial seed for SXFEL, the sliced beam energy spread is assumed to be 84 keV. And other parameters used in simulation are listed in Table 1.

Table 1. Main parameters of SXFEL operating at two-stage HGHG mode.

|  | Units | Stage-1 | Stage-2 |
|---|---|---|---|
| Seed pulse duration (FWHM) | fs | 100 | 55 |
| Seed laser wavelength | nm | 264 | 44 |
| Seed laser peak power | MW | 200 | 1000 |
| Seed laser radius (RMS) | mm | 1.3 | 0.8 |
| Modulator length | m | 1.5 | 1.5 |
| Modulator period | mm | 80 | 40 |
| Chicane dispersion $R_{56}$ | μm | 65 | 7 |
| Radiator length | m | 3.0×3 | 3.0×6 |
| Radiator period | mm | 40 | 23.5 |
| Radiator β function | m | 8-10 | 8-10 |
| Radiator wavelength | nm | 44 | 8.8 |

To obtain realistic simulation results, the whole electron beam is tracked from the first stage to the second stage. The simulation results without taking into wakefields accounts are illustrated in Fig. 6. What should be emphasized here is that FEL peak brightness evolution along the undulator is presented to include all the information from a time-dependent simulation, instead of the conventional FEL pulse energy and/or peak power. In Fig. 6, the steady increase of the peak brightness before saturation is mainly contributed by the exponential gain and the coherence build up, while the reduction of the peak brightness after saturation and during the insert modules are mainly caused by the diffraction of the FEL transverse profile. In more detail, 44 nm FEL achieves a maximum brightness of $0.5 \times 10^{29}$ photons/(mm$^2$×mrad$^2$×s×0.1%BW) at the end of the first stage radiator, and the 8.8 nm FEL shows a maximum brightness of $2.5 \times 10^{29}$ photons/(mm$^2$×mrad$^2$×s×0.1%BW) after 4 segments of the radiator.

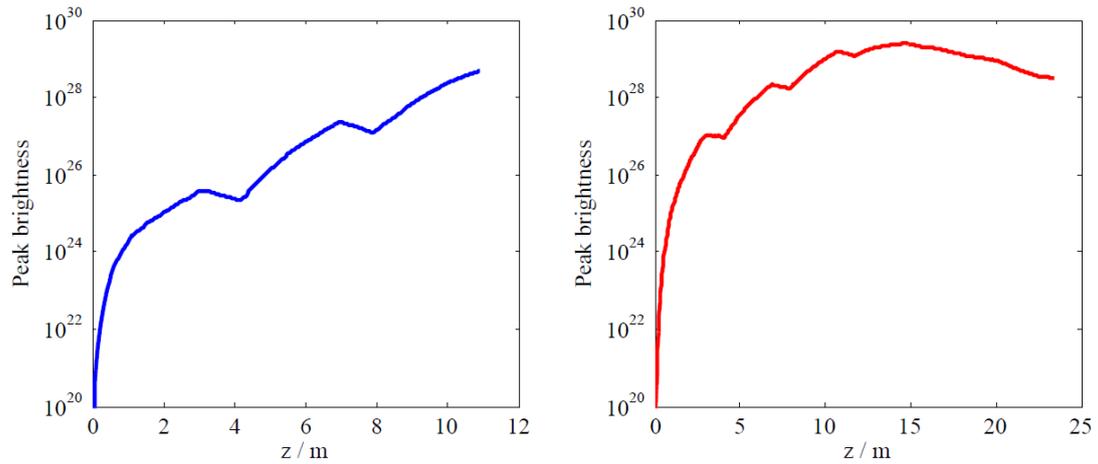

Fig. 6. Peak brightness evolutions of the 44 nm FEL (left) and 8.8 nm FEL (right). The brightness unit is photons/(mm$^2$×mrad$^2$×s×0.1%BW).

Then the FEL performances, i.e., the radiation power pulse and spectrum at the optimal position of

Fig. 6 are given. In order to obtain more reliable results, the wake effects in the undulator section have been taken into account. As shown in Fig. 7, although there is a slight degradation in the FEL power distribution in the calculation with wake effects, the generated FEL power and FEL spectrum at the first stage are almost not affected by the wakefields. Considering the large FEL pierce parameter and the small energy loss in the first stage, it is reasonable. In the second stage HGHG, a significant peak power descend from 413 MW to 160 MW is caused by the wakefields. Meanwhile, FEL bandwidth narrowing and accompanied sidebands appear for 8.8 nm FEL in the spectrum domain, because of the saddle FEL pulse in the time domain. In the presence of wakefields, the time bandwidth product of 8.8 nm FEL increases from 0.65 to 0.75, which however is still close to the Fourier Transform Limit.

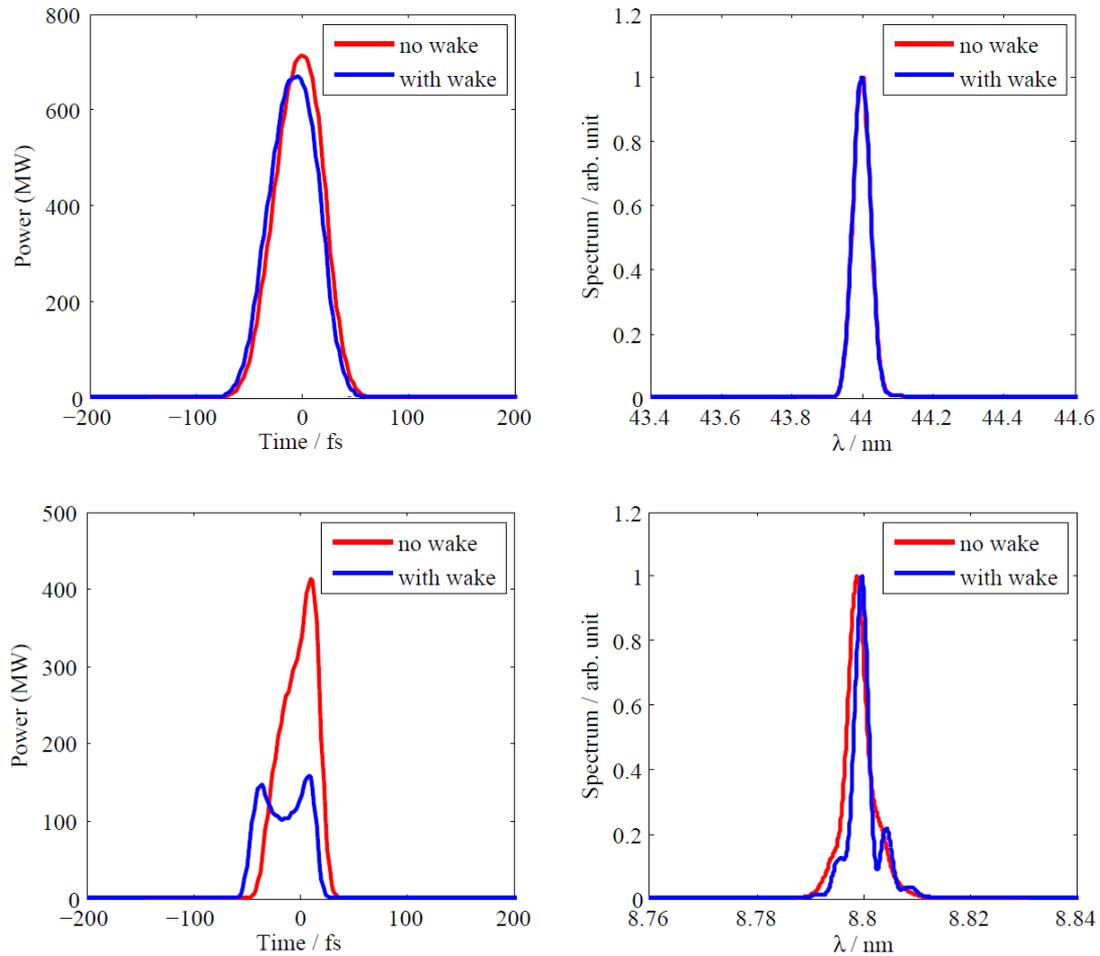

Fig. 7. The upper shows the output FEL pulse and the spectrum after 3 segments of the 44 nm radiator. The lower shows the output FEL pulse and the spectrum after 4 segments of the 8.8 nm radiator.

Fig. 8 illustrates the overall beam energy loss in the undulator section. As expected in the case without wakefields, the electron beam energy is transformed to FEL power, where the two well regions witness the interactions between the FEL pulse and the electron beam in the two stages of HGHG. However, if the wakefields are taken into accounts, there exists a large background in the beam energy loss, which makes the electrons far away from the designed FEL resonance, and results in a relatively small FEL induced energy loss, especially in the second stage. In order to compensate the energy loss caused by wakefields, one may slightly tune the magnetic field of each radiator undulator to match the FEL resonance. As an example, the dimensionless undulator parameters of the 4 segments of radiator are tuned from 1.4280 to 1.4260, 1.4255, 1.4251, and 1.4247, respectively. Then the 8.8 nm peak power reaches 320 MW again and the brightness returns to $2.5 \times 10^{29}$ photons/(mm$^2 \times$mrad$^2 \times$s$\times$0.1%BW). It is demonstrated that, the undulator fine tuning technique is of great importance for cure of wakefields induced beam energy loss. And an accuracy of 1.5 μm undulator gap control is required for the SXFEL.

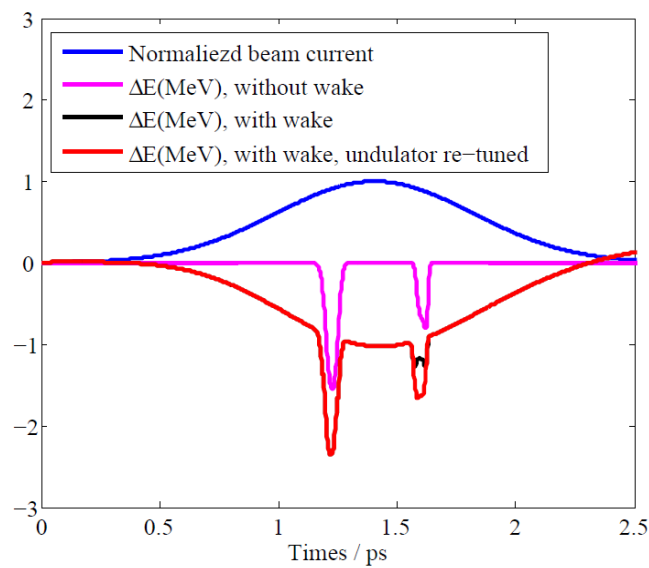

Fig. 8. The beam energy loss in the undulator section.

## 5. Conclusion

Usually, the electron bunch is relatively long and the beam pipe is regular in the LINAC section of a FEL machine, thus the wakefields generated in the LINAC is pretty smaller than that generated in the undulator section where the electron bunch is rather short and the vacuum chambers are sophisticated. Meanwhile, the beam energy loss in the LINAC section can be fixed by tuning of the RF amplitude. In this paper, the longitudinal wakefields in the SXFEL undulator section including resistive wall, surface roughness and geometrics are numerically investigated, based on the more recently technical design aspects of SXFEL. Our results demonstrate that, for a two-stage seeded FEL aiming for 8.8 nm fully coherent FEL pulses, the wakefields are not a major concern for the 1$^{st}$ stage FEL since it has a rather large FEL pierce parameter of 0.25% at 44 nm wavelength. There are definite degradations on the peak power and spectral properties for the 2$^{nd}$ stage FEL, and appropriate fine tuning of the undulator gap is useful to recover the losing performance, as verified by the simulation.

**Acknowledgements**

The authors are grateful to Xiao Hu, Meng Zhang and Bocheng Jiang for helpful discussions and useful comments. This work was partially supported the National Natural Science Foundation of China